\documentclass[journal,comsoc]{IEEEtran}
\usepackage[T1]{fontenc}
\usepackage{caption}
\usepackage{cite}
\usepackage[pdftex]{graphicx}
\usepackage{subfigure}
\usepackage{amsmath}
\usepackage[cmintegrals]{newtxmath}
\usepackage[caption=false,font=normalsize,labelfont=sf,textfont=sf]{subfig}
\usepackage{url}
\usepackage{color}
\begin{document}

\title{Two Mirroring And Interpolating Methods To Estimate Peak Position For Symmetric Signals With Single Peak}

\author{Wei Chen \\{JiangXi University of Finances and Economics}
\thanks{}
\thanks{}}

\markboth{}%
{Shell \MakeLowercase{\textit{et al.}}: Bare Demo of IEEEtran.cls for IEEE Communications Society Journals}

\maketitle

\begin{abstract}
Signals with single peak and symmetry property are very common in various fields, such as probability density function of normal distribution. Among the information contained in such signals, peak position is the most important, sometimes even the only parameter concerned. Current methods for peak position estimation are always with low precision and bad noise resistance, perform badly for sparse spectrum. This manuscript proposes two new algorithms that take advantage of symmetry property, conduct mirroring and interpolating operations to condense signal spectrum. From tests done in this paper, these two algorithms indicate outstanding advantages compared with current methods.  
\end{abstract}


\IEEEpeerreviewmaketitle

\section{Introduction}

\IEEEPARstart{S}{ymmetric} signals with single peak are widely observed in many fields. They can model amounts of phenomena, such as the distribution of white noise and reflective spectrum of a fiber Bragg grating (FBG) sensor \cite{FBGGau}. Due to their frequent appearance, the estimation for relevant parameters is urgently required and of great meaning. Several typical parameters are peak amplitude, peak position, and bandwidth. Among them, peak position is the most important, since sometimes it is difficult to determine the type of target signals, or unavoidable disturbances induce distortion to the original well-shaped waveform \cite{GauDis,GauBro}, while just symmetry is confirmed or remained. Thus peak position is the only meaningful parameter. However, current methods either focus on some specific signals and extract full parameters, like Caruana's method and Guo's method for Gaussian functions \cite{Caruana,Guo}; or fail to make full use of symmetry, thus feature with low precision and weak noise resistance, such as the centroid method \cite{Centroid}.

Currently, several methods have been applied to estimate only peak position, mainly including the centroid method \cite{Centroid}, polynomial fitting method \cite{Polyfit}, and spline fitting method \cite{Spline}. Centroid method just calculates the gravity of the selected spectrum. Polynomial fitting method views the spectrum as a polynomial function, such that the accuracy is limited. While the last algorithm uses a cubic spline to do the fitting instead, thus with better performance compared with the polynomial fitting method. Also, there are some other intelligent algorithms \cite{ant,swarm,DeepLearning}, such as the ant colony algorithm \cite{ant}, used to achieve the same purpose, but whose time cost is always too high.

This paper introduces two novel methods tended to detect the peak position for such symmetric signal with single peak. They are both basically iterative, consequently, high accuracy can be achieved. These two algorithms apply mirroring and interpolating operations to the original spectrum based on symmetry property, aiming to introduce more available samples, thus improve resistance to the noise effect. Since by nature such methods can provide a denser spectrum, they especially show good performance in processing sparse signals from our tests.



\section{Proposed new algorithms}

In some situations, for example, in temperature or strain measurement using FBG sensors, the peak position of the Gaussian-shaped spectrum is the only parameter focused on. Furthermore, in fact, due to fluctuation of source and micro interference like fiber bending, the received FBG spectrum has an imperfect Gaussian profile \cite{GauBro}, but only symmetry property maintains. It is improper to use those methods designed to fetch complete parameters like Caruana's method. So, it will be beneficial to develop a method to derive individual peak position without involving other parameters.

Currently, the centroid method is the most commonly used to solve this problem directly, as described in \eqref{centroid}. 
\begin{equation}
x_p=\frac{\sum_{i=1}^{N}x_i y_i}{\sum_{i=1}^{N}y_i} 
\label{centroid}
\end{equation}
$x_i$ and $y_i$ are the amplitude and position of the received sample, respectively. Despite simpleness in the calculation, there are some fatal drawbacks in processing such signals. One is the dilemma in the selection of points. To maintain more information from the original signal, collecting more samples spreading from the peak sounds a good idea, but meanwhile, noise problem emerges since noise would exert more effect on points with low amplitude, and vice versa. Also, it features low precision and stability as the variation of sampling rate and noise. Besides the centroid method, all other methods mentioned before have some common issues, which are the neglect of symmetry and suffering from the sparseness of the spectrum.

To overcome such disadvantages, we invented two novel algorithms so-called "Mirroring and Interpolating Methods". The newly proposed algorithms take advantage of signal symmetry, conduct interpolation to condense the spectrum, then iterate to achieve high precision. Based on such thought, we introduced two ways to get peak position only. One is similar with the centroid method while using the new spectrum obtained after mirroring and interpolating. The second one is based on the "Least Square Method (LSM)", aiming to minimize the squared error between the interpolating points and corresponding mirroring points.

First of all, a criterion for sample selection must be defined. To balance between symmetry and noise effect, acting like a band-pass filter, points are chosen bidirectionally starting from the peak position, until their amplitude lower than a defined threshold. Then filtered spectrum is sent to the next step for further processing.

\subsection{Mirroring and Interpolating Method of Type I}
The main procedure of the first new algorithm can be divided into four steps:\\
\textbf{Step 1}. In the $i^{th} $ iteration, for every sample $(x_k,y_k)$, find the corresponding  $x_k^\prime$ symmetric with respect to the $(i-1)^{th}$ peak position $ x_p^{(i-1)} $, that is 
\begin{equation}
x_k^\prime=2x_p^{(i-1)}-x_k  \label{sym_x}
\end{equation}
\textbf{Step 2}. Find the interval location $j$ for each  $x_k^\prime$, which means to search an interval $( x_j, x_{j+1} )$ such that $x_j<x_k^\prime<x_{j+1}$. Then do the interpolation for $x_k^\prime$ with two end points $(x_j,y_j)$ and $(x_{j+1},y_{j+1})$ to obtain $y_k^\prime$.\\
\textbf{Step 3}. Build the new spectrum, that is to gather both original samples  $\left\{(x_1,y_1),(x_2,y_2)\cdots(x_N,y_N)\right\}$ and interpolating points $\left\{\cdots,(x_i^\prime,y_i^\prime),\cdots\right\}$. Subsequently, calculate the $i^{th}$ peak position $x_p^i$ using the centroid method for such new spectrum
\begin{equation}
x_p^i= \frac{\sum_{i=1}^{N}(x_i y_i+x_i^\prime y_i^\prime)}{\sum_{i=1}^{N}(y_i+ y_i^\prime)}  \label{SIMT1}
\end{equation}
\textbf{Step 4}. Compare $ x_p^{i-1} $ and $x_p^i$. If $\left |x_p^i-x_p^{(i-1)}\right |$ is lower than defined threshold value, stop iteration; else back to Step 1.

\begin{figure}[!t]
	\centering
	\includegraphics[width=2.5in]{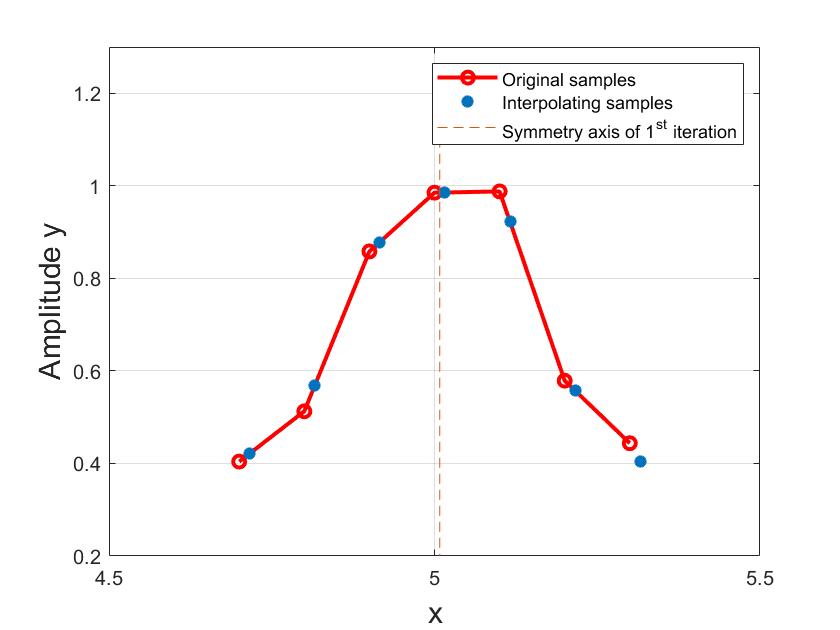}
	\caption{Mirroring and interpolating method of type I}
	\label{MIM1}
\end{figure}

The initial value $x_p^{(0)}$ can be set to the peak position of the original spectrum. About the interpolation, either linear or nonlinear is feasible. In fact, at step 2, there may be some point of this set $\left\{ \cdots, x_i^\prime, \cdots \right\}$ which will not locate in this range $(x_1,x_N)$ and thus can not be interpolated. For this condition, we just generate the respective symmetric point without interpolation instead, shown as the blue point rightmost in Fig. \ref{MIM1}. Its following procedures are performed the same as normal points.

Taking advantage of signal symmetry, this method uses mirroring and interpolating operations to condense the spectrum, thus introduces more useful points and improves the peak estimation performance.
\\
\subsection{Mirroring and Interpolating Method of Type II}
\begin{figure}[!t]
	\centering
	\includegraphics[width=2.5in]{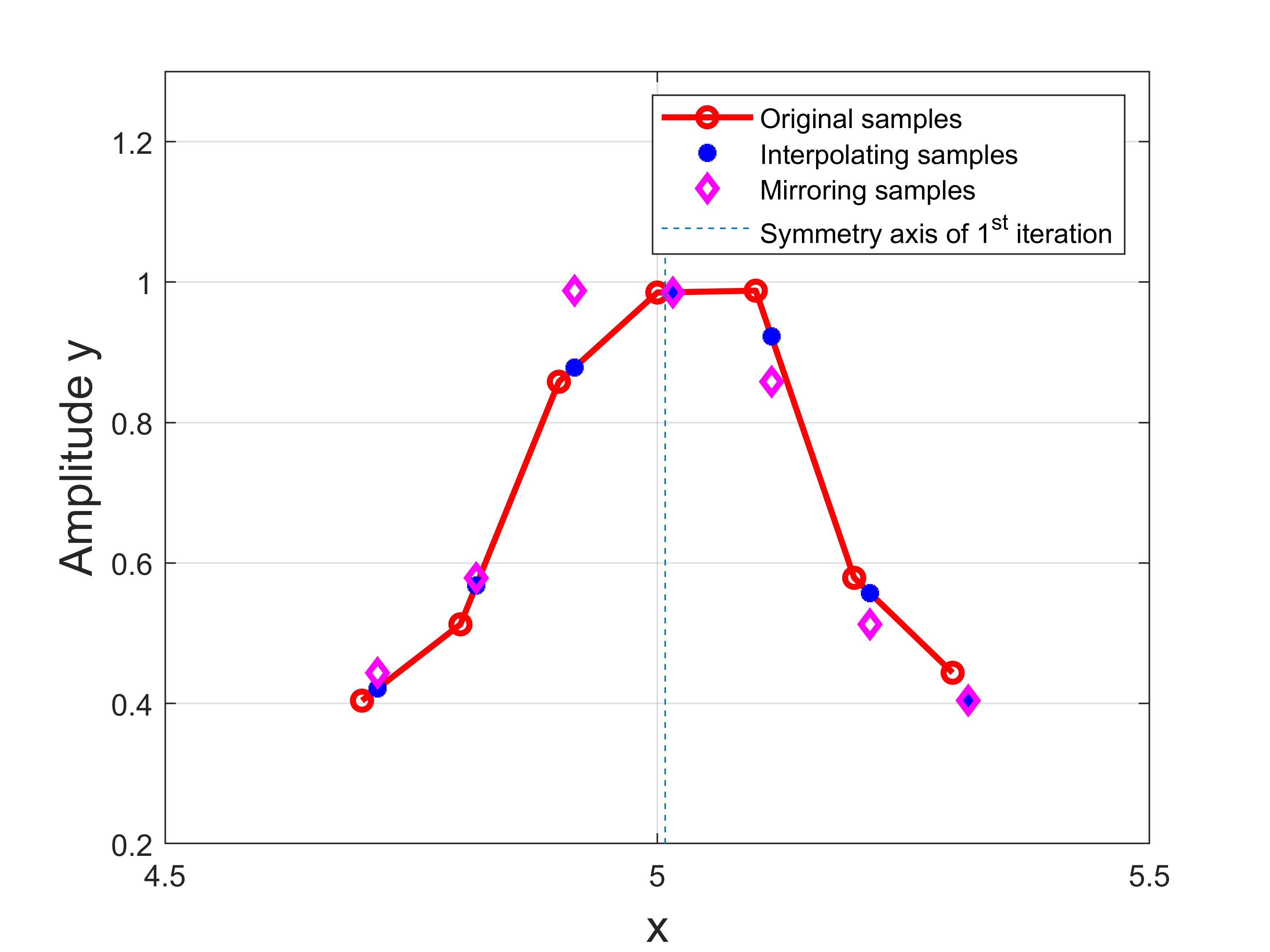}
	\caption{Mirroring and interpolating method of type II }
	\label{MIM2}
\end{figure}

This algorithm is designed to derive the peak position by minimizing the difference between mirroring points and interpolating points. The basic idea is that, since the spectrum is symmetric, the points generated by mirroring each original sample with respect to the true symmetric axis located at the peak position will perfectly belong to the spectrum. However, because the true and complete spectrum is unknown, we can do interpolation for the original samples to approximate the mirroring points. Therefore, the condensing operations for the spectrum branches -- one mirroring, and the other interpolating. Since for each time of mirroring, the symmetric axis is the current peak position, not the unknown true one, obviously, these two spectra would be different, shown as in Fig. \ref{MIM2}. Consequently, our target is to find such a peak position that can minimize the square sum of difference, which comes to a common and easy problem.\\
Without loss of generality and for convenience to obtain the explicit form, we use linear interpolation here, and note $(x_i^M,y_i^M)$ and $(x_i^I,y_i^I)$ as mirroring point and interpolating point of $(x_i,y_i)$ respectively. In fact, $y_i^M$ is equal to $y_i$, $x_i^M$ is the same as $x_i^I$. Note this square sum of difference as S, that is
\begin{equation}
S=\sum_{i=1}^N (y_i^I-y_i^M)^2 \label{prd}
\end{equation}
As mentioned in step 2 of "Mirroring and Interpolating Method of Type I" and taking linear interpolation as example, we can rewrite $y_i^I$ using the two neighbor end points$(x_j,y_j)$ and $(x_{j+1},y_{j+1})$, that is
\begin{equation}
y_i^I=\frac{y_{j+1}-y_j}{x_{j+1}-x_j}(x_i^I-x_j)+y_j \label{y_i}
\end{equation}
Since the original spectrum is uniform sampled, thus $x_{j+1}-x_j$ is constant, noted as $d_x$. Furthermore, $x_i^I$ can be expressed as $2x_p-x_i$, $x_p$ is the peak position, then \eqref{y_i} is shown as
\begin{equation}
y_i^I=\frac{y_{j+1}-y_j}{d_x}(2x_p-x_i-x_j)+y_j \label{y_i2}
\end{equation}
Then \eqref{prd} also is transferred into
\begin{equation}
S=\sum_{i=1}^N (\frac{y_{j+1}-y_j}{d_x}(2x_p-x_i-x_j)+y_j-y_i)^2 \label{fDiff}
\end{equation}
Optimizing this problem by setting the differential of  $S$ with respect to $x_p$ to zero, and the solution of such equation comes to 
\begin{equation}
x_p=\frac{\sum_{i=1}^N (\frac{(y_{j+1}-y_j)^2}{d_x^2}(x_i+x_j)+\frac{y_{j+1}-y_j}{d_x}(y_i-y_j))}{\sum_{j=1}^{N}\frac{(y_{j+1}-y_j)^2}{d_x^2}} \label{LSM}
\end{equation}
However, we can not obtain $x_p$ directly from \eqref{LSM} as the interval location $j$ corresponding to $(x_i,y_i)$ is dependent on the previous peak position. Therefore, this algorithm is also iterative by nature. Its iteration procedure is described as following:\\
\textbf{Step 1}. In $k^{th}$ iteration, for point mirrored by every $x_i$ with respect to ${(k-1)}^{th}$ peak position $x_p^{k-1}$, noted as $x_i^M$, find the interval $(x_j,x_{j+1})$ such that $x_i^M$ is contained in.\\
\textbf{Step 2}. Calculate the $k^{th}$ peak position $x_p^k$ using \eqref{LSM} based on obtained information from Step 1.\\
\textbf{Step 3}. Define stop criterion for the iteration. That is firstly to compare the difference between consecutive two results, $x_p^{k-1}$ and $x_p^k$. Then depending on the difference, end the iteration if it is lower than the threshold, otherwise back to Step 1.

In practice, occasionally, especially in a high power noise environment, the iteration would not converge but oscillate between two final values. In such a case, we can terminate the iteration by setting a limit value for the iteration number.\\

For these two new algorithms, besides the symmetry, prior information about other characters of the target signal is not mandatory. This advantage would largely broaden their applications in fitting symmetric functions. In addition, the algorithms are iterative, implying a potential high accuracy. Moreover, due to the interpolating operation in both two algorithms, a denser spectrum is formed, which indicates lower noise susceptibility. This effect is rather beneficial to the fitting of a signal with a sparse spectrum or low sampling rate. Also, they are low time cost, proper for real-time processing.

\section{Simulation and comparison}
In this section, we do simulation for three algorithms mentioned above, which are the centroid method and two types of mirroring and interpolating methods, since the three are all simple, general, and direct. The performance of these algorithms under different noise levels has been studied. To verify the capability of analyzing sparse spectrum, tests have been repeated at several sampling rates. Finally, the sample selection criterion would be modified, aiming to search the optimal set-up for the threshold value.

Firstly, to evaluate the sensitivity to noise, without loss of generality, we select a Gaussian function as an example, with amplitude $A=1$, peak position $\mu=5$, and standard deviation $\sigma=0.2$. The sampling rate is set to 10, signal interval ranges from 0 to 10 in this test. A white noise with zero mean value is added. Its noise level $\sigma_n$ is assigned with several values, from 0.2, 0.15, 0.1, 0.05, 0.025, 0.01 to 0.005, corresponding SNR equal to 14dB, 16.5dB, 20dB, 26dB, 32dB, 40dB and 46dB. Empirically, $0.5\sigma_n$ is chosen as the sample amplitude threshold as a compromise, such that about 10 points are selected. Finally, the experiment would be repeated 5000 times under each noisy environment. The results of mean value and variance are analyzed.
\begin{figure}
	\centering
	\includegraphics[width=2.5in]{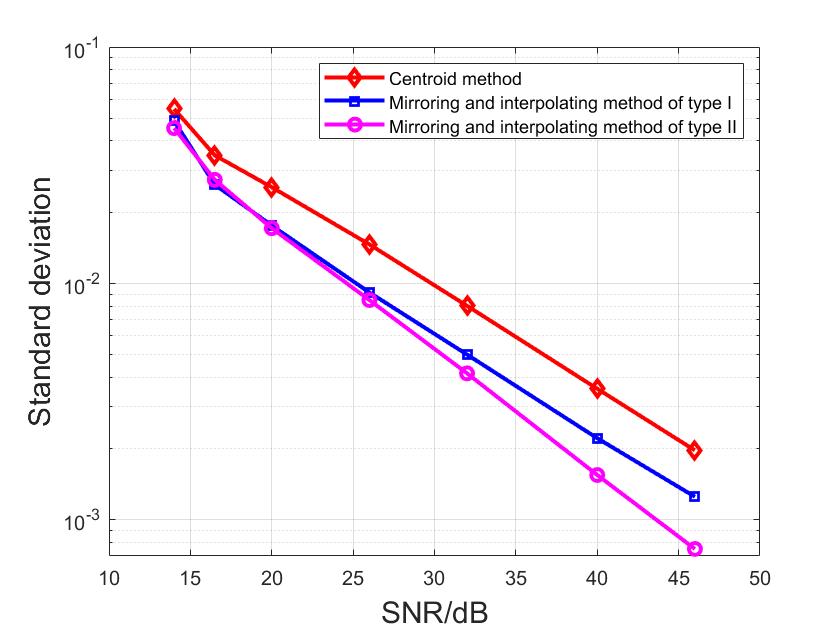}
	\caption{Precision comparison for peak position estimation at different SNR}
	\label{precisionComSNR}
\end{figure}

From both current and following tests, it is observed that the bias from the true value of all three methods is very small, always several orders lower than that of standard deviation. Precision value dominates the peak detection quality, so only the result of precision has been illustrated in this paper.

In Fig. \ref{precisionComSNR}, it can be found that two mirroring and interpolating methods show significant higher precision in peak position estimation compared with the centroid method, especially at the low noise level. For example, at SNR=46dB, the standard deviation of the centroid method is nearly tripled compared with that of the novel method of type II. The new algorithms perform similarly under low SNR conditions, while type II works better at high SNR.\\ 

\begin{figure}
	\centering
	\includegraphics[width=2.5in]{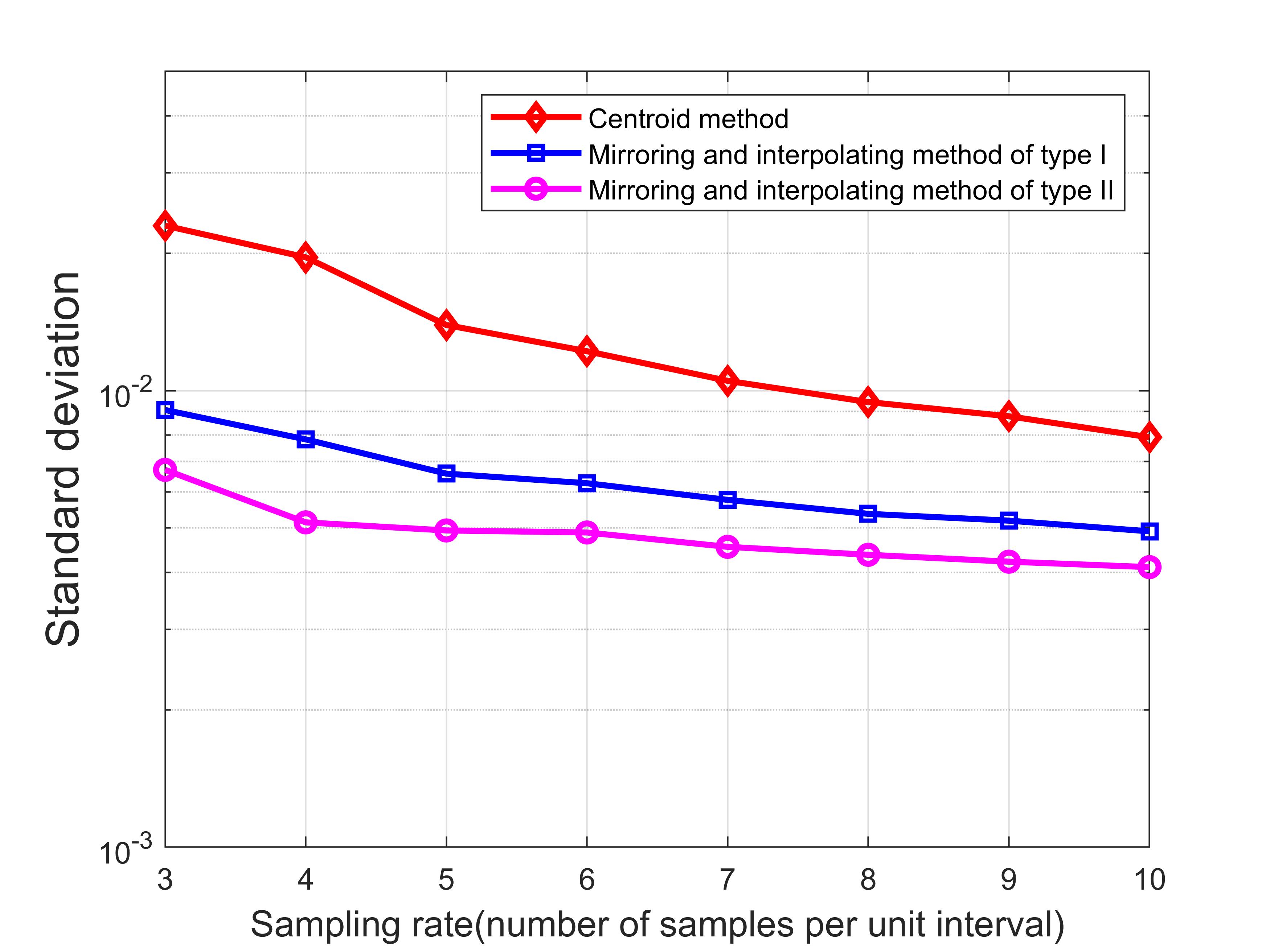}
	\caption{Precision comparison for peak position estimation at different sampling rates}
	\label{precisionComSR}
\end{figure}

The sampling rate also affects the peak detection quality, since it determines the density of the spectrum. A denser spectrum will decrease the fluctuation induced by noise since it introduces more samples. Experiments designed to investigate the influence of sampling rate have been repeated 5000 times under each sampling condition. The tested signal is also a Gaussian function with the same parameters as before. While the white noise level is fixed at 0.025, half of such value is set as the sample filtering threshold. The sampling rate has been tuned from 3 to 10, 1 at each step, the corresponding precision is shown in Fig. \ref{precisionComSR}.

From the results in Fig. \ref{precisionComSR}, the proposed two new methods perform much better than the centroid method. Among them, the new method of type II is outstanding at all sampling rates. The centroid method is the most sensitive to the sampling rate, as predicted before. However, the other two show high resistance to decreasing of sampling rate, which implies that the mirroring and interpolating approach works. As a consequence, they are very appropriate for processing signals with a sparse spectrum.

About the spectrum determination, after knowing the peak location, a common filtering procedure would be executed to its nearby samples. By moving forward and backward, samples with an amplitude higher than the threshold would be gathered to form the spectrum used for the following analysis. In such an approach, the determination of the threshold value is very important. It decides the quality of the spectrum used to process. A small threshold would let more samples with lower amplitude in, help to maintain symmetry, and suppress the noise disturbance. However, meanwhile, these samples of low power are badly affected by noise thus not correct, so it also would lower the quality of the spectrum. High threshold behaves contrarily. So there would be a trad-off for the setup of spectrum determination threshold. 

\begin{figure}
	\centering
	\includegraphics[width=2.5in]{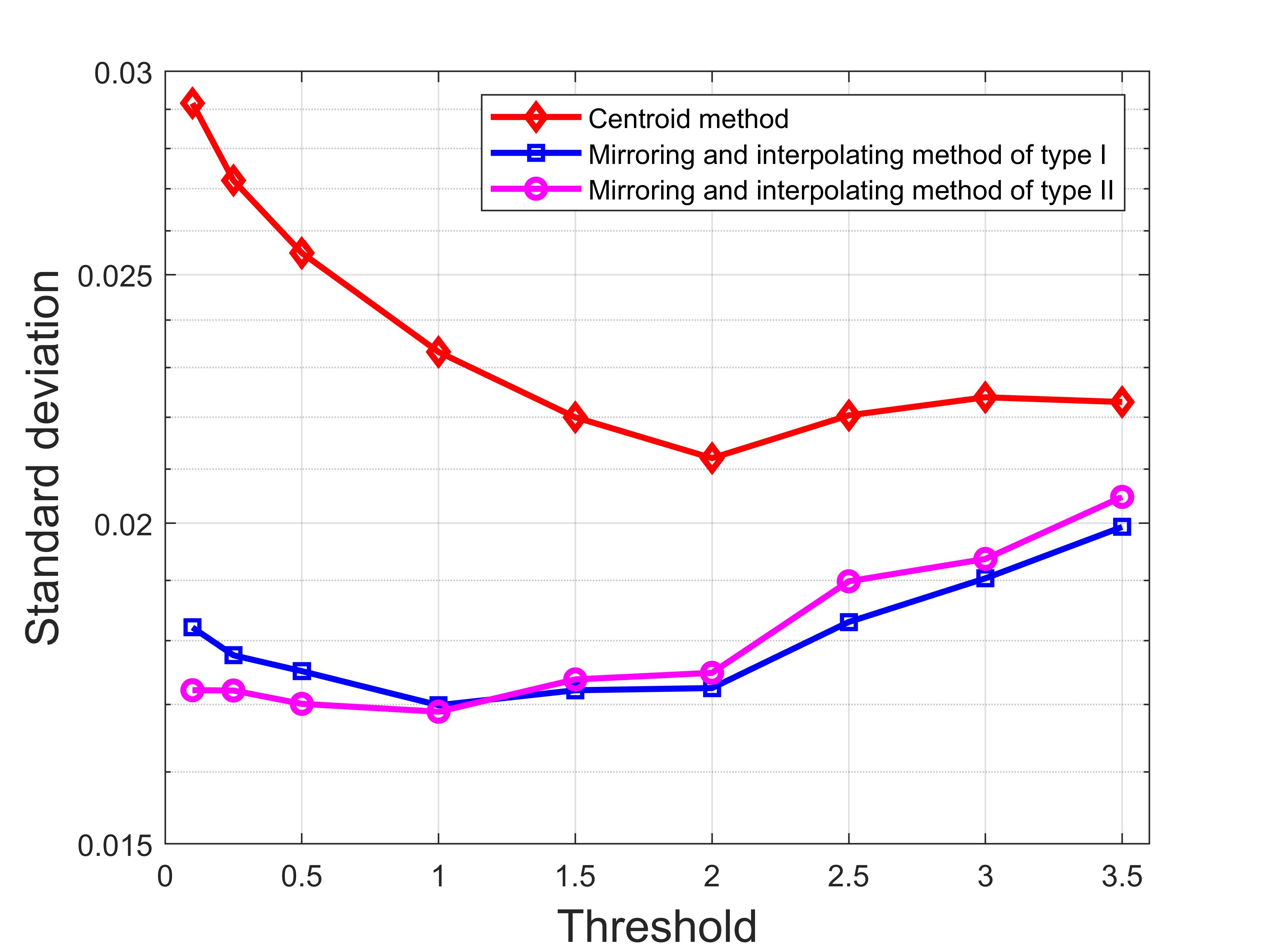}
	\caption{Precision comparison for peak position estimation at different threshold values}
	\label{precisionComTH}
\end{figure}

A test concerning the effect of such parameter has been carried out. Such threshold value is modified from 0.1$\sigma_n$ to 3.5$\sigma_n$ with SNR fixed at 20dB. Then using the same Gaussian function, peak position estimations applying the three methods have been implemented, corresponding results are exhibited in Fig. \ref{precisionComTH}. Again, precision decides the peak detection performance, accuracy comparison has been omitted. 

In Fig. \ref{precisionComTH}, it is interesting that there is an optimal threshold value for all the three methods for the Gaussian signal case, though less significant for the new methods. Such optimal value is about 2$\sigma_n$ for the centroid method, $\sigma_n$ for type I and II. From the results shown here, it can be concluded that, in general, owing to the mirroring and interpolating operation, the two novel methods provide not only a higher resistance to the introduction of points with low amplitude, but also a wider working range, which bring much convenience for real applications.

In additional, during the above simulations, the convergence for these two new methods are fast, commonly less than 10 times of iteration is required to achieve a precise result. By their nature, signal symmetry is fully and only explored. Therefore, though all tests taking a Gaussian signal as an example, naturally, these innovative methods can handle well the peak estimation issue for different types of symmetric signals with single peak, not limited to the Gaussian model.

\section{Conclusion and Discussion}

In this paper, two novel types of mirroring and interpolating methods have been proposed to evaluate only peak position for symmetric signals with single peak. They take advantage of the symmetry of target functions, increase spectrum density by mirroring and interpolating approach, then iterate to achieve higher precision. 

Type I works more like the centroid method, just calculates centroid for the new spectrum. While type II utilizes LSM to minimize the sum of the difference between interpolating points and mirroring points. The core idea of these two methods is to condense the spectrum to increase the resistance of noise interference and lower the fluctuation of the result by mirroring and interpolating operations.

Compared with the centroid method, such two innovative algorithms show outstanding merits, including much higher precision and less sensitivity to the sampling rate. These two methods are rather useful when precision and robustness are required but no further information about the target signal is given excluding symmetry. Moreover, the two types of such algorithms show less susceptibility to the low spectrum threshold and also broaden the proper working range, which means we can ease the worrying about the issue of sample selection. They are both direct, simple, and iterative in nature, also with fast convergence. Due to the condensing operation, they work well in processing sparse spectrum or when high sampling rate is unavailable.


%


\ifCLASSOPTIONcaptionsoff
  \newpage
\fi











\bibliographystyle{ieeetr}

\bibliography{ref}
\end{document}